\documentclass[reprint,
 amsmath,amssymb,
 pre,aps,
]{revtex4-2}

\usepackage{graphicx}
\usepackage{dcolumn}
\usepackage{bm}
\usepackage{comment}
\usepackage{xcolor}
\usepackage{svg}
\usepackage{xcolor}

\begin{document}

\preprint{APS/123-QED}

\title{Dynamical Chaos in a Dissipative Driven Quantum Soft Impact Oscillator}

\author{Titir Mukherjee}
 \affiliation{Department of Physical Sciences, Indian Institute of Science Education and Research Kolkata, Nadia 741 246, West Bengal, India}
\author{Arnab Acharya}%
\affiliation{Department of Physical Sciences, Indian Institute of Science Education and Research Kolkata, Nadia 741 246, West Bengal, India}%
\author{Soumitro Banerjee}
\affiliation{Department of Physical Sciences, Indian Institute of Science Education and Research Kolkata, Nadia 741 246, West Bengal, India}%
\author{Deb Shankar Ray}
\affiliation{Indian Association for the Cultivation of Science, Jadavpur, Kolkata 700 032, India}%

\date{\today}

\begin{abstract}
Dynamical chaos in a periodically driven, dissipative soft impact oscillator is investigated in the quantum regime using the complex-number quantum Langevin equation (c-number QLE). The averaged system dynamics are analyzed through a comprehensive suite of time-series diagnostics, including bifurcation diagrams, Lyapunov exponents, Fourier spectra, and the $0$--$1$ test. Systematic variation of the wall position reveals a rich sequence of dynamical transitions and grazing bifurcations, progressing from periodic to multiperiodic motion and culminating in chaotic behavior. These results demonstrate the persistence of impact-induced chaos under quantum dissipation and elucidate how environmental fluctuations influence non-linear dynamics in open quantum systems.
\end{abstract}

\maketitle


\section{Introduction}
Understanding chaos in quantum systems is a fundamental challenge, as classical notions such as trajectories and sensitivity to initial conditions do not apply directly to this realm. The study of classical chaotic systems in the quantum domain has led to the emergence of a new field known as quantum chaos. One of the most popular approaches to studying quantum chaos is through eigenvalue statistics \cite{jensen1992quantum,izrailev1990simple,stockmann2007quantum}. Another line of research uses the out-of-time-order correlator (OTOC) to understand information scrambling in quantum systems, by tracking the exponential growth of operator non-commutativity over time \cite{garcia2022out,hashimoto2017out}. However, there is limited literature applying dynamical system analysis tools to explore the average dynamics of quantum observables \cite{laha2020time,laha2020bifurcations,lakshmibala2022nonclassical}.  In this work, we explore this direction by analyzing the average dynamics of operators in a quantum dissipative system that exhibits rich non-linear phenomena classically.

Classical piecewise smooth dynamical systems \citep{nordmark1991non, bernardo2008piecewise, banerjee2009invisible} are well known for exhibiting rich and interesting behaviors. One notable example is the impact oscillator \cite{peterka2004phenomena,ing2008experimental}, which shows features such as grazing-induced bifurcation, abrupt transition to chaos, and multistability. Although classical chaotic systems such as the kicked rotor \cite{santhanam2022quantum} and billiard systems \cite{richter1999playing} have been widely explored in the quantum domain, quantum impact oscillators have received much less attention.  \cite{acharya2023signatures} reported quasiperiodic behavior in the quantum impact oscillator without forcing, and the appearance of strange nonchaotic dynamics in the system with forcing. However, no studies of the same system have been reported in the presence of dissipation. 

In the quantum regime, dissipation is usually modeled by coupling the system to an environment represented as a bath of harmonic oscillators. Using this approach, extensive studies have been conducted on decoherence and energy loss in systems such as quantum oscillators \cite{lin_quantum_dissipation} and spin chains \cite{zhu_spin_chain, zarei_spin_chain2025}. It has been shown that decoherence can also arise without energy dissipation, in what is known as dissipationless decoherence \cite{milburn_decoherence2001, sharma_decoherence2025}. 

Recent works have highlighted that dissipation, rather than simply destroying coherence, can, in fact, prolong chaotic features in certain quantum systems \cite{chaos_dissipation2025}. Quantum many-body systems, in particular, have become an active playground for investigating the interplay between chaos, entanglement, and thermalization under dissipation \cite{manybody_theory2025, manybody_experiment2024}. Various classically chaotic systems, such as the kicked top \cite{Chaudhury2009, kicked_top2}, the Duffing oscillator \cite{duffing1, duffing2}, and the Morse oscillator \cite{morse_oscillator}, have also been explored in the quantum regime with dissipation, offering insights into how classical chaos emerges in quantum settings.

Although the Lindblad formalism \cite{manzano2020short,brasil2013simple,breuer2002theory} remains the standard tool for modeling open quantum dynamics, it becomes increasingly unwieldy for continuous, nonsymmetric systems due to the need for an infinite operator expansion. The quantum Langevin equation approach can effectively tackle this problem. Starting from a system-bath model, we transition to the Heisenberg picture and eliminate the bath degrees of freedom to derive an effective set of equations governing the system's dynamics \cite{Ford1988, agarwal1971brownian}. This method provides a more tractable and physically transparent means of studying dissipative quantum systems like the impact oscillator.

In this paper, we study the dynamics of a forced soft impact oscillator in the presence of dissipation using the quantum Langevin equation framework. We demonstrate that, in the presence of environmental noise, the system exhibits a rich dynamical behavior that depends on the position of the wall. By varying the wall position, we observe bifurcations from one periodic behavior to another and from periodic to chaotic dynamics. The occurrence of each dynamical behavior is confirmed by bifurcation diagrams, FFT spectra, Lyapunov exponent analysis, and the 0-1 test. In particular, we identify parameter windows where the average dynamics shows a positive Lyapunov exponent and returns a 1 in the 0-1 test, indicating sensitive dependence on initial condition, a hallmark of chaos. This work demonstrates that tools from classical dynamical systems can effectively capture the onset of chaos in quantum dissipative systems.

The structure of this paper is designed to present a coherent progression from classical to quantum analysis of the soft impact oscillator. In Section II.A, we introduce the classical model. Section II.B outlines the derivation of the quantum Langevin equation with complex number noise. In Section II.C, we detail the numerical scheme used to solve the c-number quantum Langevin equations, including parameter choices, time-stepping techniques, and handling of discontinuities. Section III presents an analysis of quantum behavior, utilizing diagnostics such as phase-space trajectories, power spectra, and Lyapunov exponents, as well as the O-1 test, to highlight signatures of chaos. Finally, Section IV summarizes the principal findings and discusses their broader implications for the study of quantum chaos in non-smooth dynamical systems.

\section{Methodology}

\subsection{The Classical soft impact oscillator}

\begin{figure}[h]
    \centering
    \includegraphics[width=0.75\linewidth]{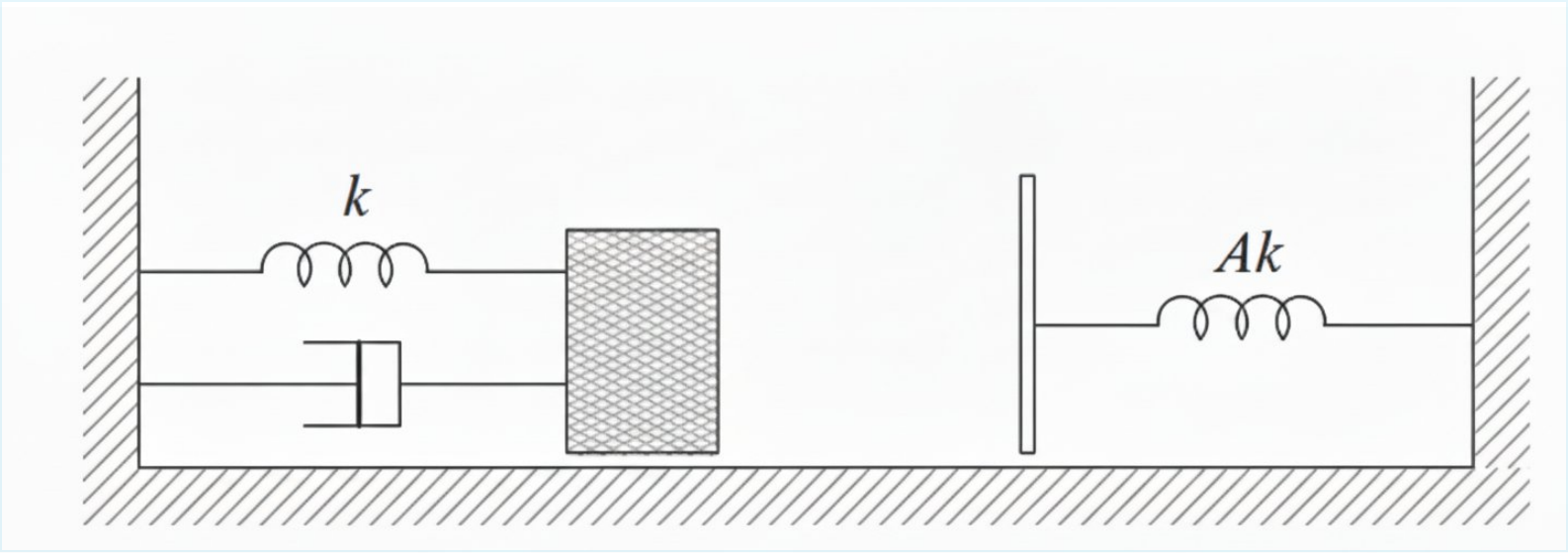}
    \caption{Schematic diagram of a soft impact oscillator. }
    \label{fig: diagram}
\end{figure}

\begin{figure}[tbh]
    \centering
    \includegraphics[width=0.95\linewidth]{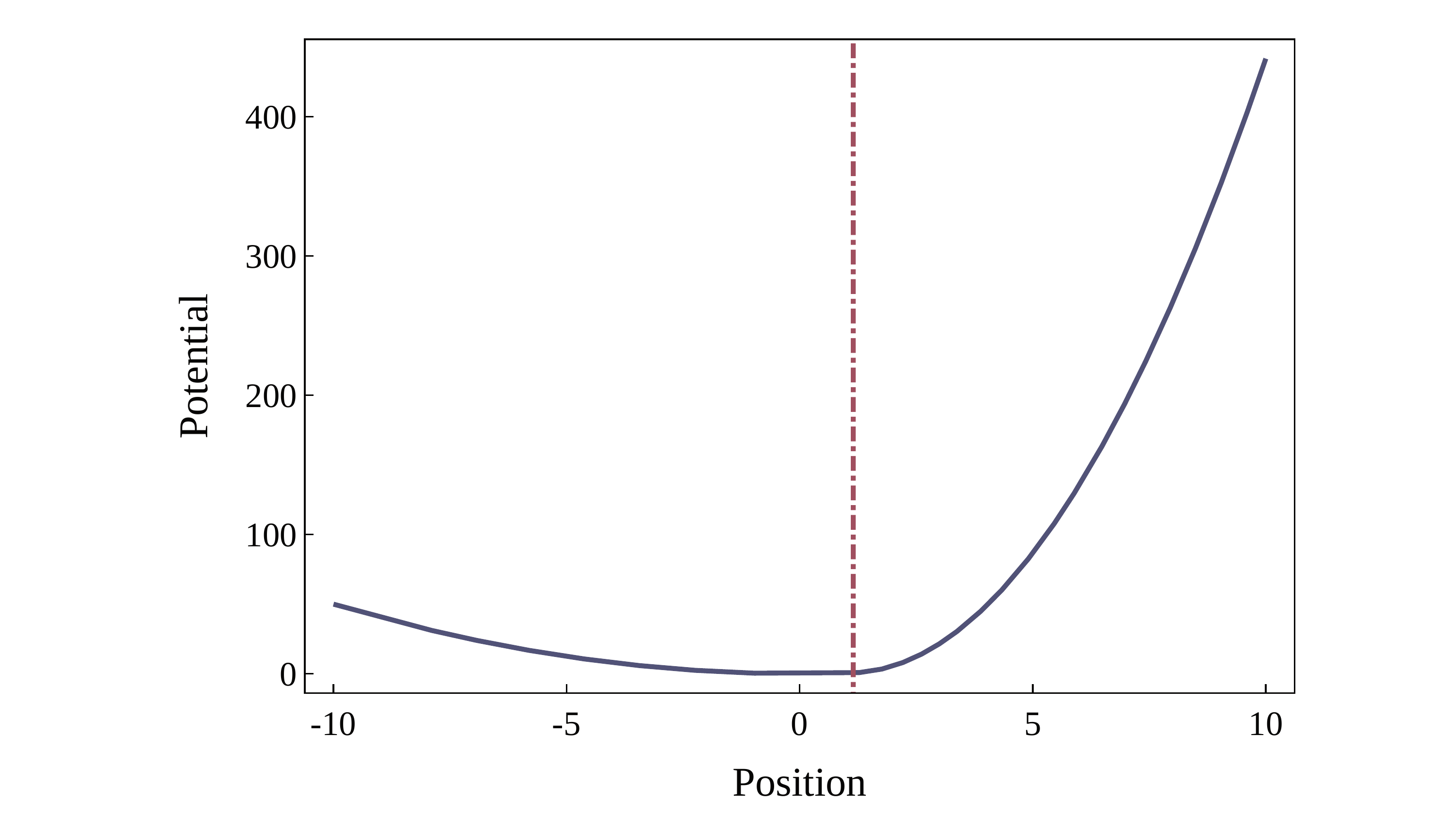}
    \caption{Graphical representation of the soft impact potential}
    \label{fig: pot_graph}
\end{figure}

An impact oscillator is a system consisting of a mass-spring system with a wall in front of it. In the case of a soft impact oscillator, the wall is cushioned by another spring (stiffer than the previous one), as shown in Fig.~\ref{fig: diagram}.
One can write the potential function for this system as follows. 
\begin{eqnarray}
&V(x)&= \frac{1}{2} k x^2\text{~    if  ~}x<x_{\text{wall}} \nonumber \\
&=& \frac{1}{2} k x^2+\frac{1}{2} A k\left(x-x_{\text{wall}}\right)^2\text{~    if  ~} x \ge x_{\text{wall}}
\label{eqn: pot}
\end{eqnarray}
The graphical representation of the potential given by (\ref{eqn: pot}) is shown in Fig.~\ref{fig: pot_graph}.

The forced impact oscillator is constructed by adding a periodic term in the potential,
\begin{equation}
    V(x,t)=V(x)+xF\cos(\Omega t)
    \label{timedependent}
\end{equation}
where $F$, $\Omega$ are the amplitude and frequency of the forcing function, respectively.

\subsection{C-number quantum Langevin equation}

In the quantum domain, the system can be visualized as a single particle moving in a potential function described by Eq.~(\ref{timedependent}). This system, with a time-dependent Hamiltonian, is coupled to an environment (bath) composed of a large number of harmonic oscillators.
Dissipation, which represents the energy loss from the system to its surroundings, arises from the interaction of the particle with its environment. 

 This framework is widely used in quantum statistical mechanics to describe open quantum systems. The total Hamiltonian, which includes both the system and the bath, can be written as
\begin{equation}
\hat{H}=\frac{\hat{p}^2}{2}+V(\hat{x},t)+\sum_j\left\{\frac{\hat{p}_j^2}{2}+\frac{1}{2} \kappa_j\left(\hat{x}_j-\hat{x}\right)^2\right\}.
\label{eqn: QLE}
\end{equation}
Here, $\hat{x}$ and $\hat{p}$ denote the position and momentum operators of the system, while $\hat{x}_j$ and $\hat{p}_j$ represent those of the $j$-th bath oscillator, respectively. We take the mass of the particle $m=1$ for simplicity. The coupling constants $\kappa_j$ characterize the strength of the bilinear coupling between the system and each bath mode, and $\omega_j$ is the frequency of the $j$-th bath oscillator. All operators are considered in the Heisenberg picture unless otherwise stated. 

Following the standard derivation procedure \cite{Ford1988,agarwal1971brownian}, we write the equations of motion for both the system and the bath and then eliminate the bath variables. Physically, this means that we only track the motion of the system while the influence of the bath appears through a damping term (memory) and a fluctuating term (noise). This leads to the operator-based quantum Langevin equation (QLE):
\begin{equation}
\ddot{\hat{x}}(t)+\int_0^t d t^{\prime} \gamma\left(t-t^{\prime}\right) \dot{\hat{x}}\left(t^{\prime}\right)+V^{\prime}(\hat{x},t)=\hat{F}(t).
\label{eqn: OQLE}
\end{equation}

The first term represents acceleration, the second term represents friction or dissipation due to the environment, and the last term on the right, $\hat{F}(t)$, is the fluctuating force exerted by the environment on the system. The time integration appearing in Eq.~(\ref{eqn: OQLE}) embodies the system's memory, which means that the present dynamics of the system depend on its past. The function $\gamma(t)$, called the memory kernel, quantifies how this dependence fades with time:
\begin{equation}
\gamma(t) = \sum_j \kappa_j \cos(\omega_j t).
\label{eqn: memory}
\end{equation}

In simpler terms, each bath oscillator contributes an oscillating term to the overall damping, and when summed over many oscillators, interference between different frequencies causes $\gamma(t)$ to decay over time. This decay represents how quickly the system forgets its past. In the Markovian limit, when the bath does not have memory, $\gamma(t)$ reduces to a Dirac delta function.

The random force or noise operator $\hat{F}(t)$ is given by
\begin{equation}
\hat{F}(t)=\sum_j\left[\left\{\hat{x}_j(0)-\hat{x}(0)\right\} \kappa_j \cos \omega_j t + \hat{p}_j(0) \kappa_j^{1 / 2} \sin \omega_j t\right]
\end{equation}
This expression shows that the noise originates from the initial fluctuations of the bath variables, since the bath is assumed to be in a thermal state at the initial time. These microscopic fluctuations are propagated forward in time through the oscillatory functions $\cos(\omega_j t)$ and $\sin(\omega_j t)$, which encode the free evolution of each bath oscillator. Consequently, even though bath degrees of freedom are not explicitly simulated, their random influence on the system persists through the noise term $\hat{F}(t)$.

Numerically solving Eq.~(\ref{eqn: OQLE}) directly is non-trivial because it involves operator-valued quantities. To make progress, we take the quantum mechanical average of this equation, which allows us to describe the evolution of the system in terms of expectation values rather than operators. In this approach, the system and bath are assumed to start from a separable initial state of the form
$|\Psi(0)\rangle = |\phi\rangle \otimes \left\{|\alpha_1\rangle|\alpha_2\rangle \cdots |\alpha_N\rangle\right\},$
where $|\phi\rangle$ denotes any initial quantum state of the system, and each $|\alpha_j\rangle$ is a coherent state representing the $j$-th harmonic oscillator of the bath. The explicit form of such a state for a harmonic oscillator is given by
\begin{equation}
|\alpha_j\rangle=\exp \left(-\frac{\left|\alpha_j\right|^2}{2}\right) 
\sum_{n_j=0}^{\infty}\frac{\alpha_i^{n_j}}{\sqrt{n_{j}!}}|n_j\rangle,
\end{equation}
where $|n_j\rangle$ is the $n$-th energy eigenstate of the $j$-th oscillator. 

Taking the expectation value of Eq.~(\ref{eqn: OQLE}) over this separable state effectively removes the operator nature of the equation and yields
\begin{equation}
\langle\ddot{\hat{x}}(t)\rangle + \int_0^t d t^{\prime} \gamma\left(t-t^{\prime}\right) \left\langle\dot{\hat{x}}\left(t^{\prime}\right)\right\rangle + \left\langle V^{\prime}(\hat{x},t)\right\rangle = f(t),
\label{eqn: aQLE}
\end{equation}
where $f(t)=\langle \hat{F}(t)\rangle$ represents the quantum mechanical mean of the noise operator. This $f(t)$ is no longer an operator but a complex-valued quantity (often called a c-number noise). 

To describe $f(t)$ statistically, we must specify how the bath oscillators are initially distributed. Physically, we assume that the environment is in thermal equilibrium at temperature $T$. The corresponding distribution of the initial positions and momenta of each oscillator is given by the Wigner distribution function of a shifted harmonic oscillator \cite{barik2003numerical,banerjee2004solution}:
\begin{equation}
P_j = N \exp \left\{-\frac{\left[\left\langle\hat{p}_j(0)\right\rangle^2+\kappa_j\left\{\left\langle\hat{x}_j(0)\right\rangle-\langle\hat{x}(0)\rangle\right\}^2\right]}{2 \hbar \omega_j\left(\bar{n}_j+\frac{1}{2}\right)}\right\}.
\end{equation}
Here, $\bar{n}_j=1 /\!\left[\exp \left(\hbar \omega_j / k T\right)-1\right]$ represents the average number of thermal excitations (or photons) in the $j$-th oscillator, and $N$ is a normalization constant ensuring that the total probability integrates to one. 

This probability distribution allows us to perform an ensemble average over many possible initial conditions of the bath. For any observable quantity $O_j$, the statistical average over the ensemble is defined as
\begin{equation}
\left\langle O_j\right\rangle_s=\int O_j P_j \, d\left\langle\hat{p}_j(0)\right\rangle \, d\left\{\left\langle\hat{x}_j(0)\right\rangle-\langle\hat{x}(0)\rangle\right\}.
\end{equation}
In this way, both quantum fluctuations and thermal effects of the environment are embedded within the noise term $f(t)$.
From this definition, we can show that~\cite{barik2003numerical,banerjee2004solution}
\begin{equation}
\langle f(t)\rangle_s = 0,
\end{equation}
\begin{equation}
\left\langle f(t) f\left(t^{\prime}\right)\right\rangle_s =
\frac{1}{2} \sum_j \kappa_j \hbar \omega_j 
\left(\operatorname{coth} \frac{\hbar \omega_j}{2 k T}\right)
\cos \omega_j\left(t - t^{\prime}\right).
\label{eqn: co}
\end{equation}
These relations show that $f(t)$ behaves as a noise term with zero average value and a finite-time correlation. 
The correlation function has the same oscillatory structure as the memory kernel $\gamma(t)$, indicating that both the noise and the dissipation originate from the same system-bath interaction. 

Finally, to account for the effects of non-linear system potentials, a quantum correction term $Q(t)$ is introduced. 
This leads to the c-number Quantum Langevin Equation (QLE):
\begin{equation}
\ddot{X}(t) + V^{\prime}(X,t) 
+ \int_0^t d t^{\prime} \, \gamma\left(t - t^{\prime}\right) 
\dot{X}\left(t^{\prime}\right)
= f(t) + Q(t),
\end{equation}
where $\langle\hat{x}(t)\rangle = X(t)$ and $\langle\hat{p}(t)\rangle = P(t)$ are the quantum-mechanical expectation values of position and momentum of the system, respectively.

\begin{figure*}[t]
    \centering
    \includegraphics[width=0.99\linewidth]{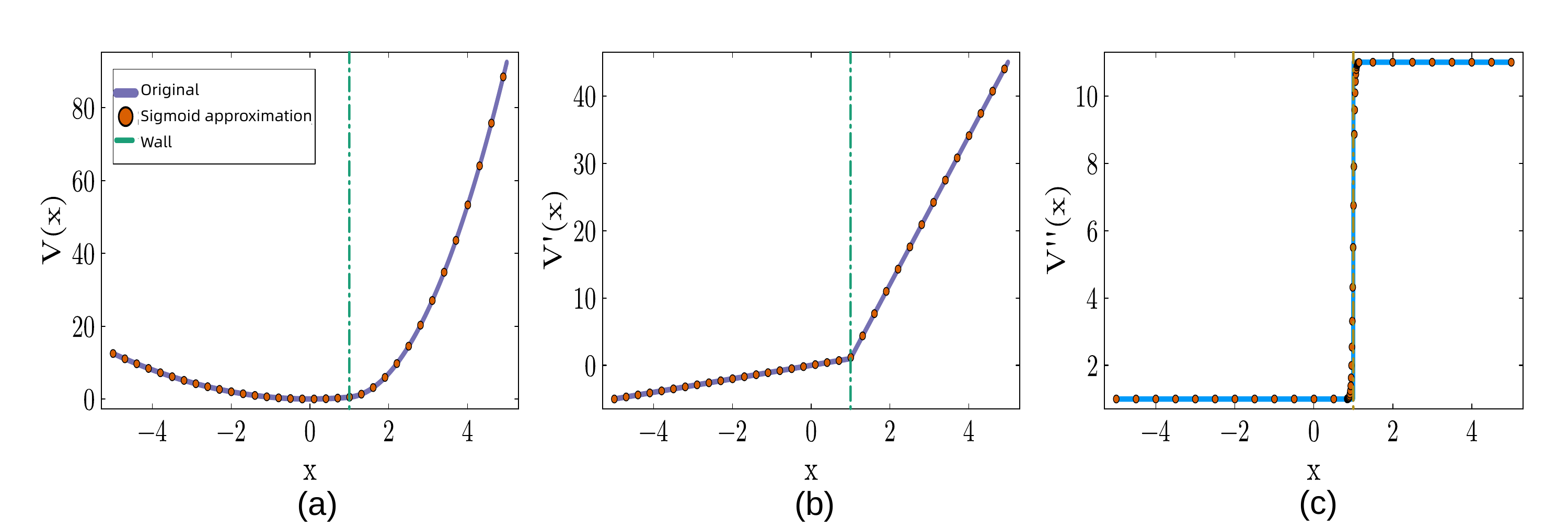}
    \caption{(a) Potential, (b) first-order derivative of potential, i.e., force on the particle, and (c) second-order derivative of potential, solid lines represent the original form, and the dots represent the form derived from sigmoid approximation}
    \label{fig: pot_d}
\end{figure*}

The correction term
\begin{equation}
Q(t) = V^{\prime}(X,t) - \left\langle V^{\prime}(\hat{x},t)\right\rangle
\end{equation}
accounts for deviations due to the non-linearity of the system in the quantum regime. 
Expanding $V'(x,t)$ around $X(t)$ gives
\begin{equation}
Q(t) = -\sum_{n \geqslant 2} 
\frac{1}{n!} V^{(n+1)}(X,t) 
\left\langle \delta \hat{x}^n(t) \right\rangle,
\label{eqn:Q}
\end{equation}
where $V^{(n+1)}$ represents the $(n+1)$th order derivative of $V$ wrt $x$, $\delta \hat{x}(t) = \hat{x}(t) - X(t)$ represents the fluctuation of the position operator around its mean value, and $[\delta \hat{x}(t), \delta \hat{p}(t)] = i\hbar$ and $\left\langle \delta \hat{x}^n(t) \right\rangle$ represents $n$-th order correction (fluctuation).

In summary, the c-number Quantum Langevin Equation provides a clear and practical way to describe the dynamics of continuous open quantum systems. 
It translates the complex operator equations into stochastic differential equations involving ordinary (complex number) variables, while still retaining essential quantum effects through the correction term $Q(t)$. This formulation makes it possible to study dissipation, noise, and quantum fluctuations within a unified and computationally accessible framework, without explicitly solving for the full density matrix or energy eigenstates.
\begin{figure*}[htbp!]
\centering
\includegraphics[width=0.3\textwidth]{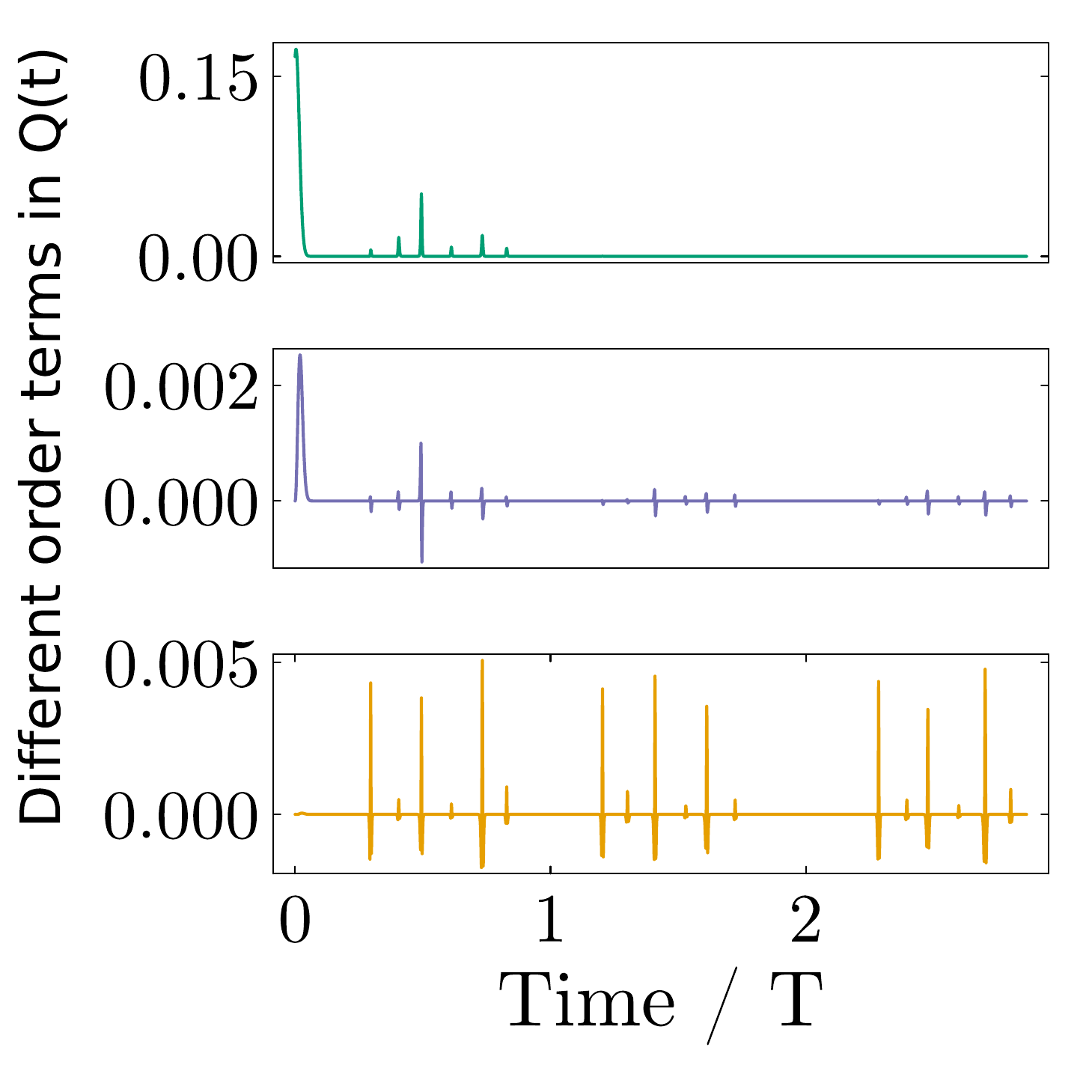}
\includegraphics[width=0.3\textwidth]{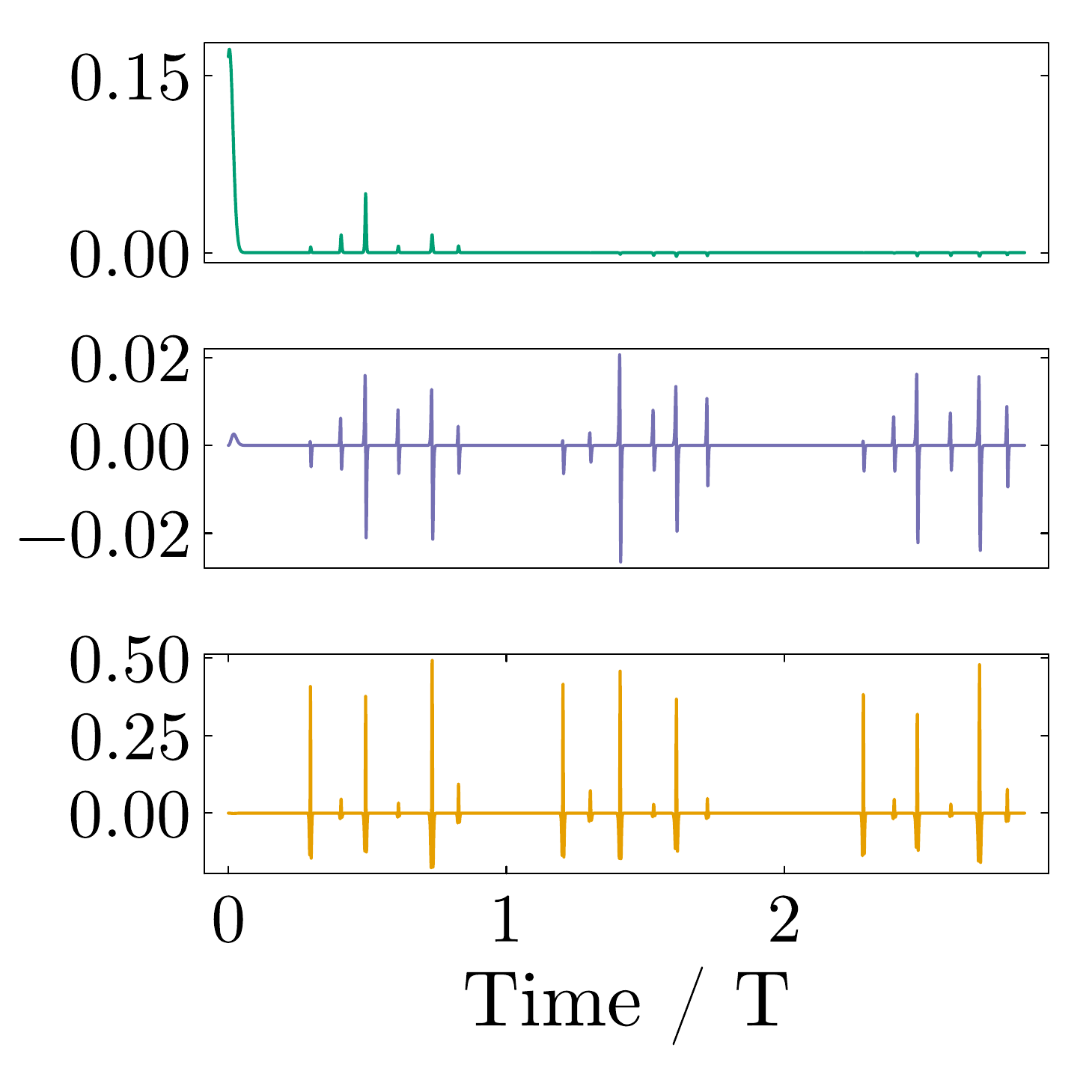}
\includegraphics[width=0.3\textwidth]{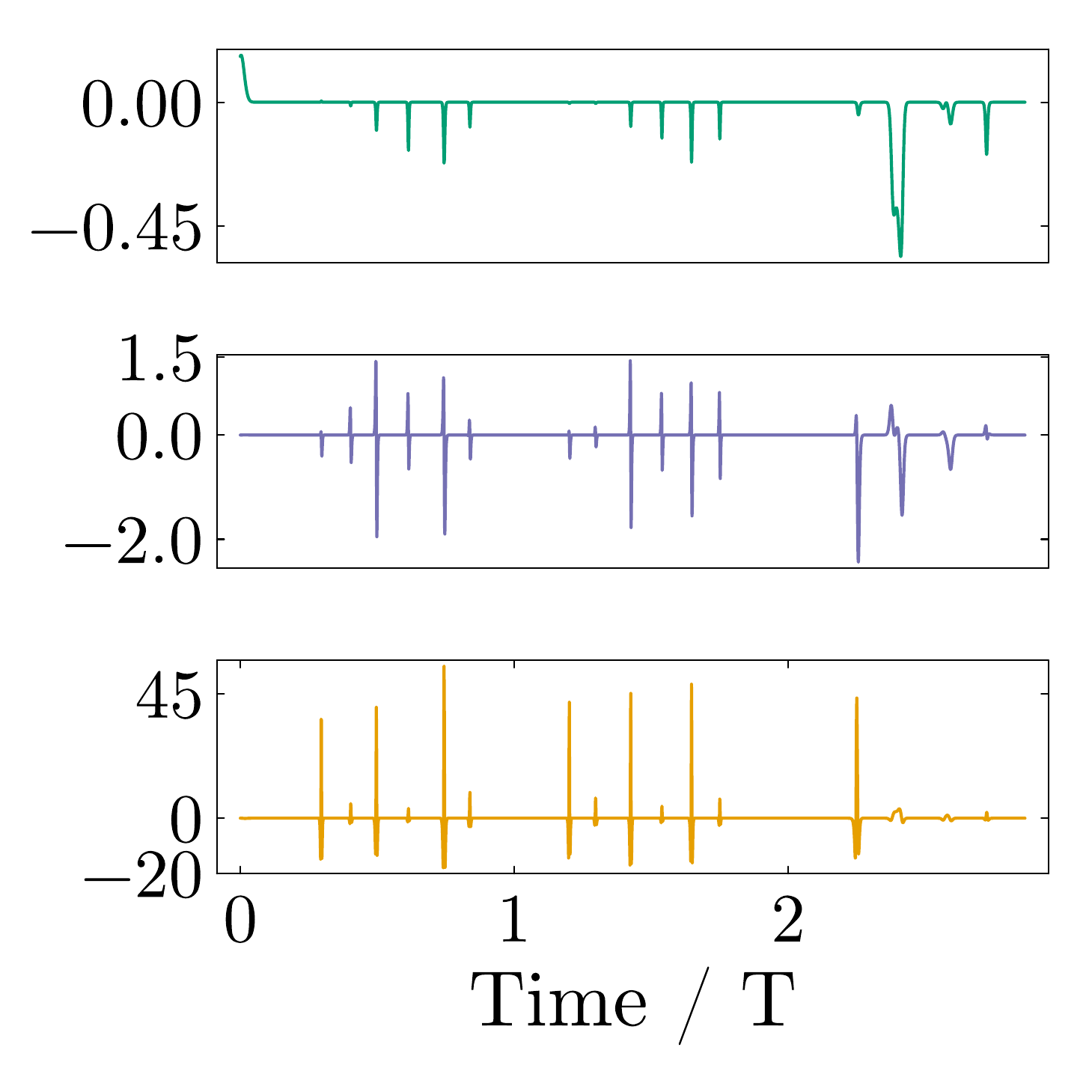}\\
{\footnotesize\hspace{0.5cm} (a) \hspace{5cm} (b) \hspace{5cm} (c)}
\caption{
Different order contributions to the quantum correction term $Q(t)$ for (a) $\hbar = 0.01$, (b) $\hbar = 0.1$, and (c) $\hbar = 1.0$, evaluated at $x_{\text{wall}} = 0.5$ and $\Omega = 0.5$.
In each column, the top, middle, and bottom panels correspond to the second, third, and fourth-order correction terms, respectively.}
\label{fig: corr} 
\end{figure*}
\subsection{Numerical simulation}
To solve Eq.~(\ref{eqn: aQLE}) numerically, we need an explicit form of the noise term $f(t)$. 
Since the environment (bath) is much larger than the system, we replace the discrete sum over bath oscillators in Eqs.~(\ref{eqn: memory}) and (\ref{eqn: co}) by an integral over frequencies:
\begin{align}
\small
\gamma(t) & = \int_0^{\infty} \kappa(\omega) \rho(\omega) \cos (\omega t) \, d\omega, \label{eqn: dis}\\
\left\langle f(t) f\left(t^{\prime}\right)\right\rangle_s 
& = \frac{1}{2}\! \int_0^{\infty}\!\!\!\!d\omega \, \kappa(\omega) \rho(\omega) \, \hbar \omega 
\operatorname{coth} \frac{\hbar \omega}{2 k T} \cos \omega(t-t^{\prime}) \nonumber \\
& \equiv c(t\!-\!t^{\prime}). \label{eqn: cor}
\end{align}

Here, $\kappa(\omega)\rho(\omega)$ is the spectral density function, which determines how strongly the system interacts with each bath frequency. Following~\cite{barik2003numerical}, we take a Lorentzian form:
\begin{equation}
\kappa(\omega)\rho(\omega) = \frac{2}{\pi} \frac{\Gamma}{1 + \omega^2 \tau_c^2},
\end{equation}
where $\Gamma$ represents the dissipation strength of the system due to the bath, quantifying how strongly the system loses energy to the environment (larger $\Gamma$ corresponds to stronger damping). 
The parameter $\tau_c$ is the correlation time of the bath,
a small $\tau_c$ corresponds to a nearly memoryless (Markovian) bath, while a large $\tau_c$ indicates a bath with long memory.
This choice of the spectral density leads to an exponentially decaying memory kernel,
\begin{equation}
\gamma(t) = \frac{\Gamma}{\tau_c} e^{-t/\tau_c}.
\end{equation}

Using the Lorentzian spectral density, the correlation function $c(t-t^{\prime})$ of the noise can be computed for given values of $\Gamma$, $\tau_c$, and temperature $kT$. 
Since $c(t-t')$ decays in time, it can be approximated as a sum of exponentially correlated noises~\cite{banerjee2004solution,barik2003numerical}:
\begin{equation}
c(t-t') = \sum_i \frac{D_i}{\tau_i} \exp \left(-\frac{|t-t'|}{\tau_i}\right), \quad i = 1,2,3,\cdots
\end{equation}
where $D_i$ and $\tau_i$ are the strength and correlation time of the $i$-th colored noise component. 
This allows us to represent the c-number noise as a sum of auxiliary stochastic variables $\eta_i$ satisfying
\begin{align}
\dot{\eta}_i &= -\frac{\eta_i}{\tau_i} + \frac{1}{\tau_i} \xi_i(t),\\
\langle \xi_i(t) \rangle &= 0, \quad 
\langle \xi_i(0) \xi_j(\tau) \rangle = 2 D_i \delta_{ij} \delta(\tau),
\end{align}
where $\xi_i(t)$ is Gaussian white noise. 
The total c-number noise is then
\begin{equation}
f(t) = \sum_{i=1}^{n} \eta_i.
\label{eqn: noise}
\end{equation}

Finally, the c-number quantum Langevin equation can be written as a set of ordinary differential equations:
\begin{equation}
\begin{aligned}
\dot{X} &= P,\\
\dot{P} &= -V^{\prime}(X,t) + Q(t) + \sum_{i=1}^{n} \eta_i + z,\\
\dot{z} &= -\Gamma \frac{P}{\tau_c} - \frac{z}{\tau_c},\\
\dot{\eta}_i &= -\frac{\eta_i}{\tau_i} + \frac{1}{\tau_i} \xi_i(t),
\label{eqn: cQLE}
\end{aligned}
\end{equation}
where the auxiliary variable $z$ arises from rewriting the integro-differential memory term as an ODE due to the exponential kernel. 
Here, $n=3$ colored noise components were used to generate $f(t)$ in practice.

For non-linear potentials, the quantum correction term $Q(t)$ includes derivatives of all orders of the potential (see Eq.\ref{eqn:Q}). However, for the soft impact potential, the second derivative is discontinuous at the wall position $x_{\text{wall}}$, making higher-order derivatives infinite. 
To avoid divergences, we approximate the second derivative with a smooth sigmoid function:
\begin{equation}
V''(x) = \frac{k A}{1 + \exp\left[-c (x - x_{\text{wall}})\right]} + k,
\end{equation}
with a large slope $c = 10$. 
By integrating this function, we reconstruct the first derivative and potential with proper constants. 
Figure~\ref{fig: pot_d} shows the quality of this approximation.

\begin{figure*}[t]
\includegraphics[width=0.9\textwidth]{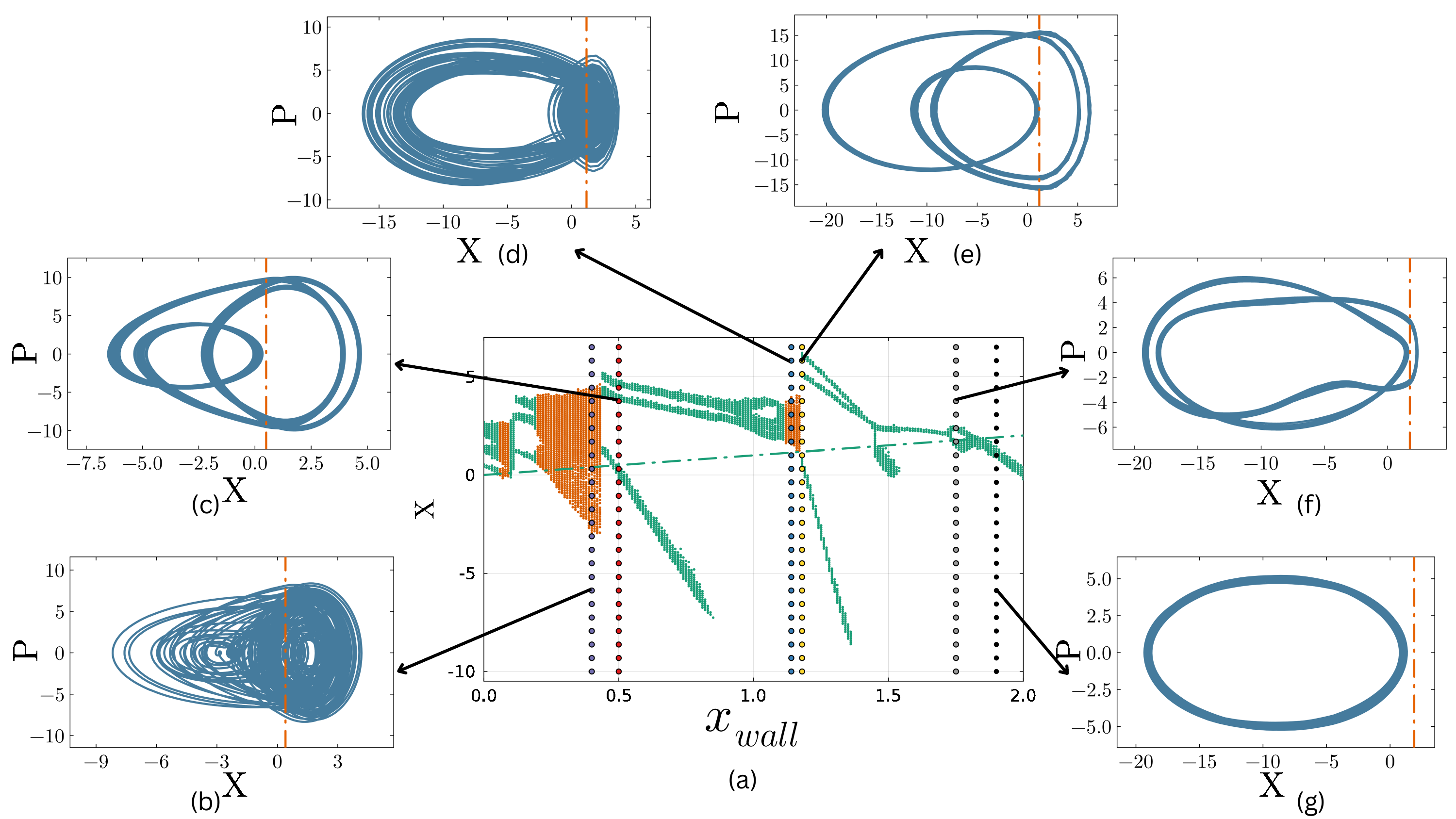}
\caption{(a) Bifurcation diagram obtained by varying the wall position $x_{\text{wall}}$ for $\Omega = 0.5$, using the Poincaré section at $P = 0$.
The color indicates the Lyapunov exponent, where green tones correspond to $\lambda \le 0$ (regular motion) and orange tones to $\lambda > 0$ (chaotic motion).
Panels~(b)–(g) show representative trajectories at selected wall positions, marked in (a) by vertical lines with circular markers:
(b)~chaotic trajectory at $x_{\text{wall}} = 0.40$ (purple),
(c)~period-three motion at $x_{\text{wall}} = 0.50$ (red),
(d)~chaotic trajectory at $x_{\text{wall}} = 1.14$ (blue),
(e)~period-three motion at $x_{\text{wall}} = 1.18$ (yellow),
(f)~period-two motion at $x_{\text{wall}} = 1.75$ (gray), and
(g)~periodic trajectory at $x_{\text{wall}} = 1.90$ (black).}
\label{fig:bif} 
\end{figure*}

\section{Results}
Earlier studies on classical impact oscillators introduced friction or dissipation and reported an abrupt transition to chaos at grazing~\cite{nordmark1991non,bernardo2008piecewise}. In the case of a soft, forced impact oscillator with dissipation, classical analyses have shown the emergence of narrow-band chaos near grazing when the forcing frequency $\Omega$ and the natural frequency $\omega$ satisfy the relation $\tfrac{2\omega}{\Omega} \neq \text{integer}$~\cite{banerjee2009invisible,kundu2012singularities}. In this work, we extend these studies to the quantum domain by investigating the dissipative dynamics of a particle in a periodically forced soft-impact potential.

\begin{figure}[tbh]
    \centering
    \includegraphics[width=0.4\textwidth]{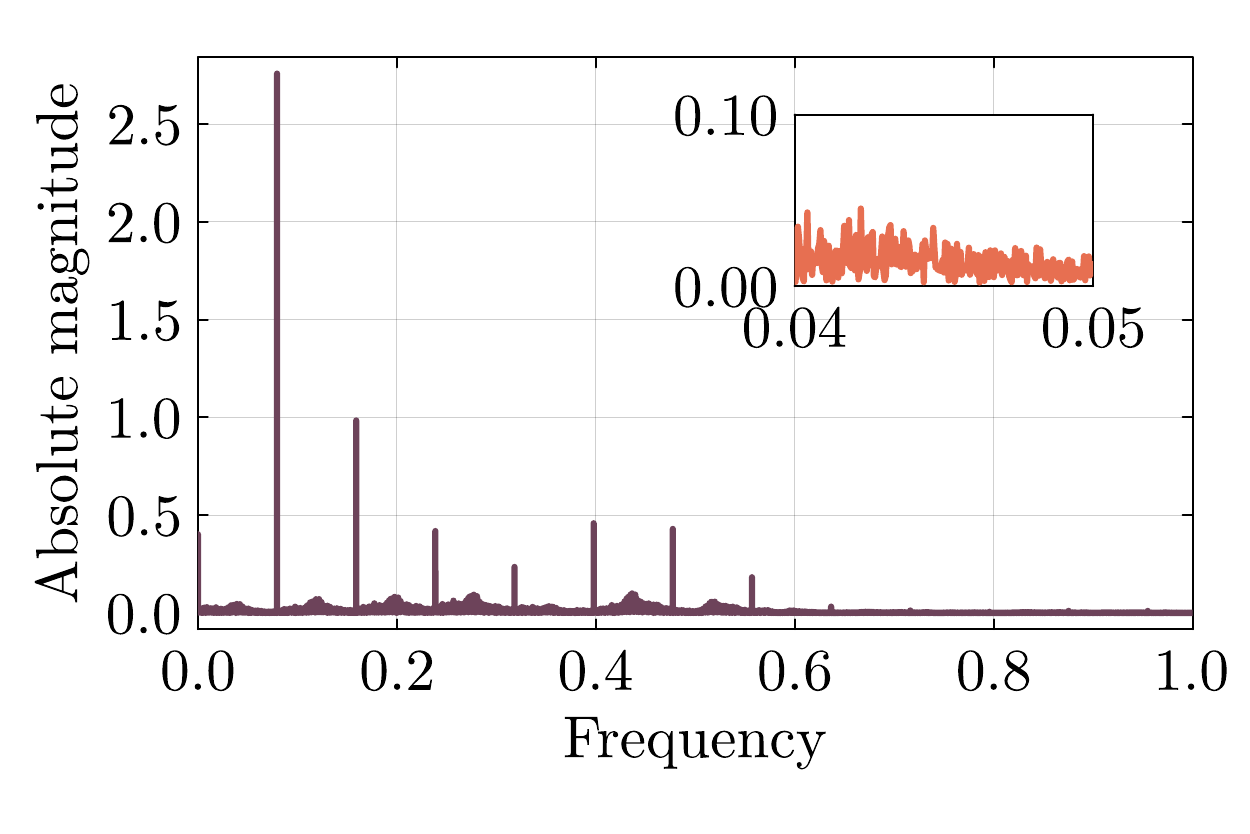}
    \\[-0.3em]
    {\centering \footnotesize (a)\\}
    \includegraphics[width=0.4\textwidth]{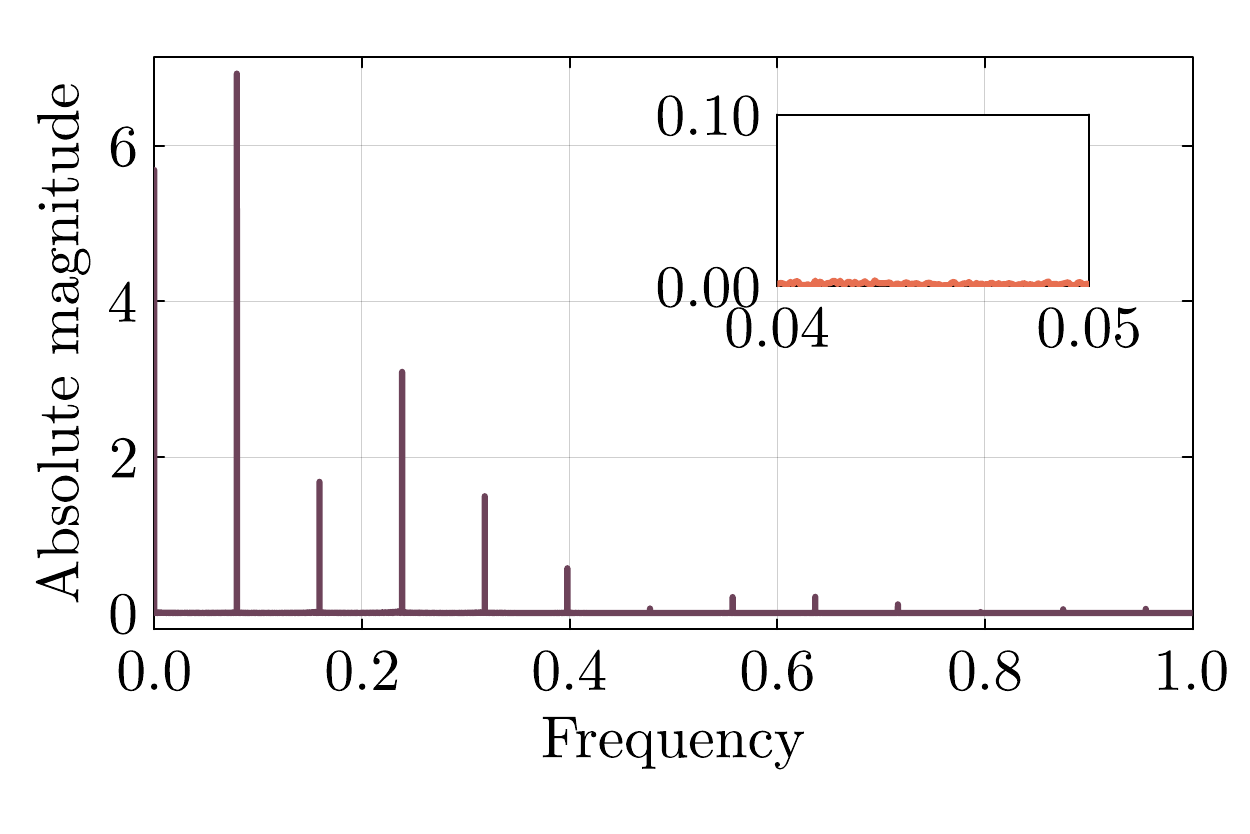}
    \\[-0.3em]
    {\centering \footnotesize (b)\\}
     \caption{FFT of the time series of $X$: (a) $x_{\text{wall}}=0.4$ showing a continuous spectrum, (b) $x_{\text{wall}}=1.35$ showing discrete peaks.}
\label{fig: FFT}
\end{figure}

\begin{figure}[tbh]
    \centering
    \includegraphics[width=0.4\textwidth]{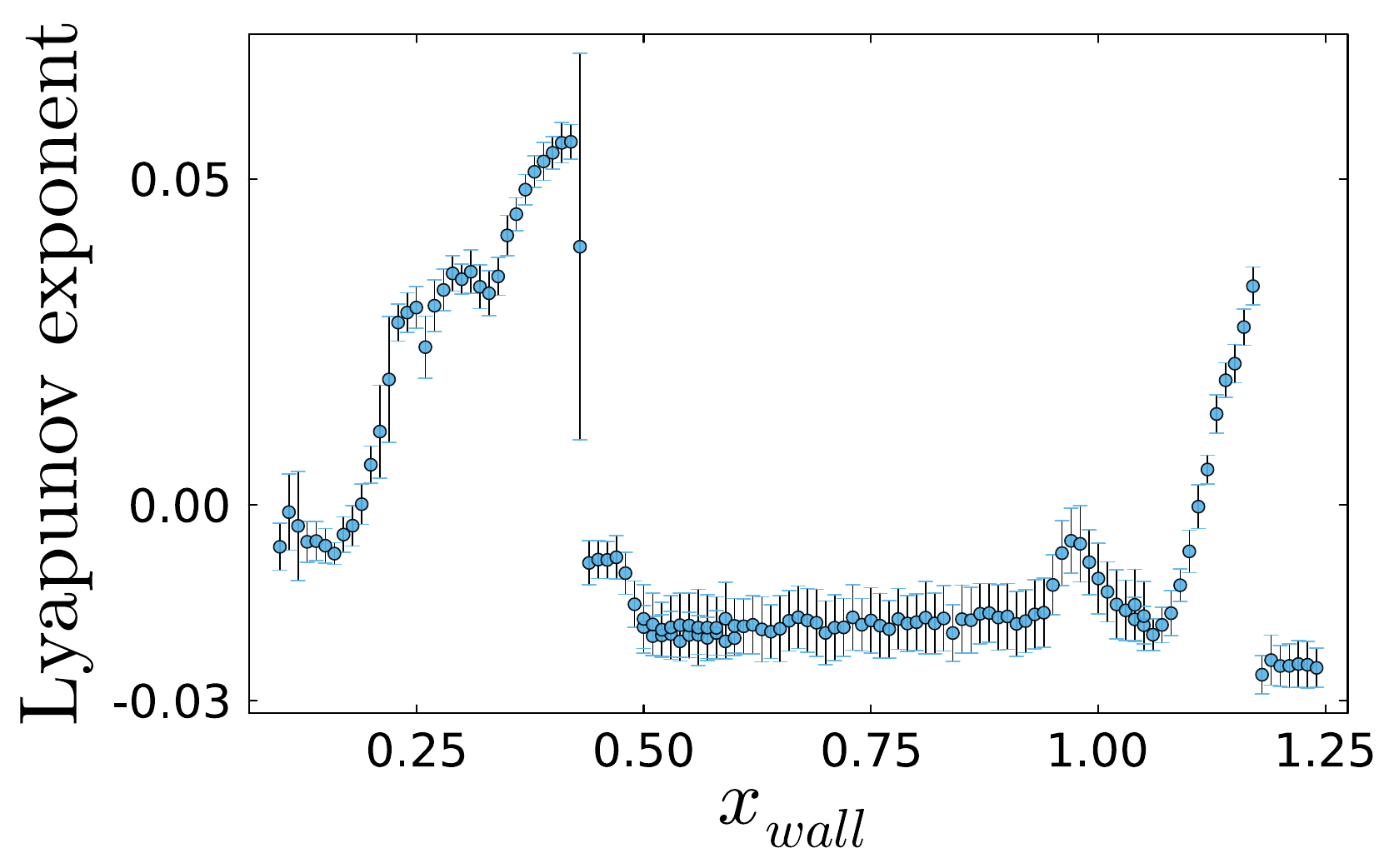}    
    \\[-0.05em]
    {\centering \footnotesize (a)\\}
    \includegraphics[width=0.4\textwidth]{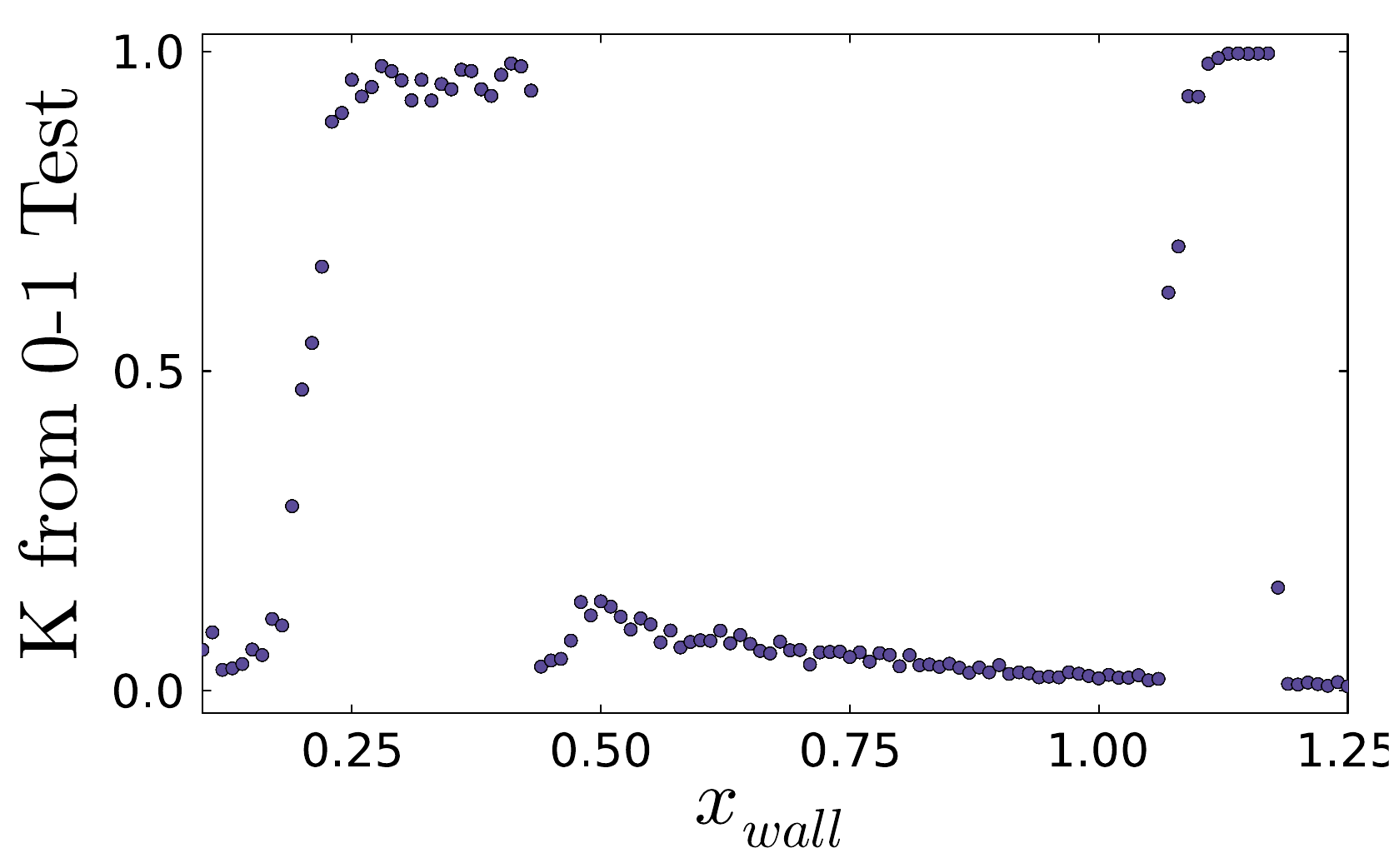}
    \\[-0.05em]
    {\centering \footnotesize (b)\\}
     \caption{(a) Mean Lyapunov exponent (sky blue) with standard deviation as error bars, computed over 1,000 noise realizations for each $x_{\text{wall}}$.  
    (b) Results of the 0–1 test as $x_{\text{wall}}$ is varied (rounded to three decimal places).}
\label{fig: measures}
\end{figure}

The parameters related to the system include the spring constant $k$, the stiffness multiplier for the stiffer spring $A$, the particle mass $m$, the position of the wall $x_{\text{wall}}$, the force amplitude $F$, and the force frequency $\Omega$. The other input parameters related to the environment are temperature $kT$, noise correlation strength $\Gamma$, and correlation time $\tau_c$. The parameters are fixed as
$k = 1.0, \quad A = 10.0, \quad m = 1.0, \quad \omega =1.0, \quad \Omega = 0.5, \quad kT = 0.01, \quad \Gamma = 1.0, \quad \tau_c = 3.0, \quad F=10$. The choice of these particular parameters makes $\tfrac{2\omega}{\Omega}=4.0$. 

We included up to the fourth-order correction to the position operator while evaluating the quantum correction term $Q(t)$. The convergence of this series was verified for each parameter regime, since the inclusion of higher-order terms beyond the fourth order becomes computationally expensive and prone to numerical instabilities. 

Figure~\ref{fig: corr} shows the time evolution of the second, third, and fourth-order contributions to $Q(t)$ for different values of $\hbar$, with $x_{\text{wall}} = 0.5$ and other parameter values as mentioned above. For $\hbar = 0.01$, the second-order term dominates with an amplitude of the order of $10^{-1}$, while the third- and fourth-order terms remain negligible, approximately $10^{-3}$, indicating rapid convergence of the series. As $\hbar$ increases to $0.1$, the magnitude of the third-order term grows to the order of $10^{-2}$, while the fourth-order term also becomes more noticeable (of the order of $10^{-1}$). However, when $\hbar$ increases to $1.0$, the relative amplitude of the fourth-order term exceeds that of the second- and third-order terms and approaches values as large as $10^{1}$.

The strong growth of higher-order terms at large $\hbar$ arises from the piecewise-smooth nature of the soft-impact potential: its second derivative exhibits a discontinuity (although we smoothed it using a sigmoid function) near the wall position $x_{\text{wall}}$, causing the third and higher derivatives to become increasingly ill-behaved. Consequently, beyond a certain threshold $\hbar$, the quantum correction series no longer converges up to the fourth order.

As the stiffness multiplier $A$ increases, approaching the hard-impact limit, this divergence appears at progressively smaller values of $\hbar$. A similar effect is observed when increasing $c$: for sufficiently large $c$ (which measures how stiff the sigmoid is), divergence occurs already at the fourth order, even for $\hbar = 0.01$. Therefore, in this work, we restrict our analysis to the semi-quantum regime with $\hbar = 0.01$ and $c = 10.0$.

The time scale of the dynamics in this paper is reported with respect to the forcing time scale, that is, $T = 12.56$~s. To eliminate transient effects, the initial $1000$ forcing cycles are excluded from the analysis, and the system dynamics are subsequently examined over the following $3000$ periods. To analyze the dynamics, we plot the bifurcation diagram using $x_{\text{wall}}$ as the variable parameter. Although our system is multidimensional, the two most significant dynamical variables, $X$ and $P$, represent the system's average position and momentum, respectively. For constructing the bifurcation diagram, we chose the plane $P = 0.0$ as the Poincaré section. Each time the trajectory crosses this plane from one side, the corresponding value of $X$ is saved and plotted as the $y$ coordinate, while $x_{\text{wall}}$ serves as the $x$ coordinate. The equilibrium position of the mass is taken to be $x = 0$. We perform all calculations using Julia~\cite{Julia-2017}, particularly employing the \texttt{DynamicalSystems.jl} module to calculate the Lyapunov exponent~\cite{Datseris2018}.

The averaged quantum bifurcation diagram is shown in Fig.~\ref{fig:bif}(a). On the extreme right of the diagram, where the wall is positioned far from the oscillator, the motion remains purely periodic, similar to that of a harmonic oscillator. A representative trajectory for this case is shown in Fig.~\ref{fig:bif}(g) for $x_{\text{wall}} = 1.90$ (black marker in Fig.~\ref{fig:bif}(a)). As the wall moves slightly left around $x_{\text{wall}} = 1.75$ [see Fig.~\ref{fig:bif}(f), gray marker], the particle intermittently interacts with the wall, resulting in a grazing event that leads to a transition from a period-one to a period-two motion.

When the wall is further moved to  $x_{\text{wall}} \approx 1.18$, the trajectory undergoes another grazing bifurcation, transitioning from a period-three motion, as shown in Fig.~\ref{fig:bif}(e) (yellow marker), to a chaotic trajectory (shown for $x_{\text{wall}} = 1.14$, Fig.~\ref{fig:bif}(d), blue marker). Moving the wall closer to the equilibrium position, around $x_{\text{wall}} = 0.50$, results in a distinct period-three trajectory (Fig.~\ref{fig:bif}(c), red marker). When the wall is shifted further to the left ($x_{\text{wall}} = 0.40$), the motion becomes completely chaotic, as illustrated in Fig.~\ref{fig:bif}(b) (purple marker), where the largest Lyapunov exponent attains a positive value.

To verify the onset of chaos, we computed the frequency spectra (FFT) for two representative cases. The FFT for the chaotic trajectory at $x_{\text{wall}} = 0.40$ is shown in Fig.~\ref{fig: FFT}(a), while the corresponding FFT for the period-3 trajectory at $x_{\text{wall}} = 1.35$ is shown in Fig.~\ref{fig: FFT}(b). As expected, the chaotic regime exhibits a continuous broadband spectrum (highlighted in the inset), whereas the periodic motion produces distinct discrete peaks, confirming the transition from regular to chaotic dynamics.

As shown earlier in Fig.~\ref{fig:bif}(a), regions with positive Lyapunov exponents are observed for $x_{\text{wall}} = [0.23, 0.43]$ and $[1.11, 1.17]$. However, the c-number quantum Langevin equation (QLE) inherently includes stochastic noise terms. To account for this stochasticity, we computed the Lyapunov exponent over 1,000 independent noise realizations for each parameter value and evaluated the mean and standard deviation, following the methodology adopted in \cite{ruidas2024semiclassical,bagnoli2006synchronization}. As shown in Fig.~\ref{fig: measures}(a), the average Lyapunov exponent remains positive for the regions mentioned above, $x_{\text{wall}} = [0.23, 0.43]$ and $[1.11, 1.17]$, and its magnitude significantly exceeds the standard deviation. This confirms that the system exhibits sensitive dependence on the initial condition, a defining feature of chaos.

Another widely used method to analyze chaotic time series is the 0–1 test~\cite{gottwald2004new}. This test takes a sampled data set as input and outputs a single value, $K$, between 0 and 1, where $K\approx 0$ indicates periodic or quasiperiodic motion and $K\approx 1$ indicates chaos. A key advantage of this method is that it does not require prior information about the system. We applied this test to our average time series and computed the corresponding $K$ values. As shown in Fig.~\ref{fig: measures}(b), for $x_{\text{wall}}$ in the range $0.23–0.43$, $1.11$–$1.17$, $K$ saturates near 1, confirming the presence of chaotic behavior in this region.

Although classical studies report the occurrence of narrow-band chaos only when $\tfrac{2\omega}{\Omega}$ has a non-integer value, our results demonstrate that in the quantum regime, chaos can emerge even for integer ratios. The value of $\tfrac{2\omega}{\Omega}$ used in our analysis is 4.0, which is an integer. We also tested different values of $\Omega$, such as $\Omega = 0.6$ (corresponding to $\tfrac{2\omega}{\Omega} = 3.33$) and $\Omega = 1.62$ (the golden ratio case), and in all these cases we observed qualitatively similar dynamical behavior—coexistence of different periodic trajectories, grazing events, and the onset of chaos.

\section{Conclusion}
In this work, we have explored the dissipative dynamics of a periodically driven soft impact oscillator in the quantum regime using the c-number quantum Langevin equation. Quantum corrections to the position operator $Q(t)$ were incorporated up to the fourth order, and the analysis was restricted to the semi-quantum limit ($\hbar = 0.01$) to ensure numerical convergence and physical consistency.

By systematically varying the wall position $x_{\text{wall}}$, we observed a rich sequence of dynamical transitions reminiscent of classical grazing bifurcations. The motion evolved from purely periodic to multiperiodic and eventually chaotic as the wall approached the oscillator. These transitions were clearly reflected in the averaged bifurcation diagram and corroborated through complementary diagnostics—namely, the Lyapunov exponent, FFT spectra, and the 0–1 test for chaos. A robust chaotic regime emerged in the range $x_{\text{wall}} = 0.23$–$0.43$, where the mean Lyapunov exponent remained significantly positive and exceeded its standard deviation.

Taken together, these findings confirm that grazing-induced chaos persists even in the semi-quantum domain, although its strength and robustness are modulated by quantum noise and dissipation. The present study thus bridges the gap between classical and quantum impact dynamics and provides a systematic framework for understanding how environmental fluctuations influence the onset and nature of chaos in open quantum non-linear systems.

From an experimental standpoint, the present results suggest two promising directions for verification.

Atomic Force Microscopy (AFM) cantilevers operated in intermittent-contact or tapping mode constitute natural realizations of impact oscillators~\cite{vanDeWater2000, Raman2006}. When driven near resonance, the cantilever intermittently interacts with the substrate, exhibiting non-linear transitions and grazing-type bifurcations. As AFM probes reach smaller tip masses and operate at cryogenic temperatures, quantum fluctuations begin to compete with thermal noise, necessitating a quantum mechanical description~\cite{Ali2017, Passian2017}. Under such conditions, the predicted quantum corrections to $Q(t)$ and the associated bifurcation signatures may become experimentally observable in next-generation quantum-limited AFM setups.

Cavity optomechanical systems governed by quantum Langevin equations~\cite{Ludwig2008} offer a complementary and highly tunable platform. Non-linear dynamics and chaos have already been demonstrated in nanomechanical and optical cavity systems~\cite{NavarroUrrios2017, Lu2022}. Furthermore, recent advances in levitated optomechanics allow for the engineering of arbitrary potential landscapes~\cite{Rakhubovsky2021}, including non-smooth, cubic, or higher-order non-linear potentials. Such developments open a pathway for experimentally probing the quantum manifestations of grazing bifurcations and chaos predicted by the present model.

Overall, this framework provides a tractable route for exploring the onset of chaos in open quantum systems with discontinuous nonlinearities, while identifying feasible experimental avenues for its realization and verification.

\section{Acknowledgements}
S Banerjee and A. Acharya acknowledge financial support
from J. C. Bose Grant of the Anusandhan National Research Foundation, Govt. of India, No.
JBR/2020/000049. T Mukherjee acknowledges financial support from  UGC, Govt. of India.
\bibliography{Quantum}

\end{document}